\documentclass[11pt,a4paper]{article}
\pdfoutput=1
\usepackage{jheppub}

\usepackage{pifont}
\usepackage{graphicx}
\usepackage{verbatim}
\usepackage{slashed}
\newcommand{\ba}{\begin{eqnarray}}
\newcommand{\ea}{\end{eqnarray}}
\newcommand{\no}{\nonumber}
\newcommand{\be}{\begin{equation}}
\newcommand{\ee}{\end{equation}}
\newcommand{\bea}{\begin{eqnarray}}
\newcommand{\eea}{\end{eqnarray}}

\title{
Spontaneous CP violation and the strong CP problem
}
\date{\today
}
\author{Luca Vecchi
}
\affiliation{Maryland Center for Fundamental Physics,\\ Department of Physics, University of Maryland\\
College Park, MD 20742, USA}
\emailAdd{vecchi@umd.edu}
\abstract{
We derive sufficient conditions that guarantee a robust solution of the strong CP problem in theories with spontaneous CP violation, and introduce a class of models satisfying these requirements. In the simplest scenarios the dominant contribution to the topological angle arises at 3-loop order in the Yukawa couplings. A variety of realizations are possible on a warped extra dimension, which can simultaneously address the Planck-TeV hierarchy. Experimental signatures of this approach to the strong CP problem include flavor violation and vector-like partners of the top or bottom quarks.
}
\begin{document}
\maketitle
%%%%%%%%%%%%%%%%%%%%%%%%%%%%%%%%%%%%%%%%%%%%%%%%%%%%%%%%%%%%%%%%%%%
%
%%%%%%%%%%%%%%%%%%%%%%%%%%

\section{Motivation}

The colored sector of the standard model (SM) has two CP-odd parameters of phenomenological interest, parametrized by two nearly RG-invariant phases. One is the CKM phase $\theta_{\rm CKM}$, defined in terms of the Jarlskog invariant. The second is the topological angle $\bar\theta=\theta_{\rm QCD}-\theta_{\rm F}$, with $\theta_{\rm QCD}$ the coefficient of $\frac{g_s^2}{32\pi^2}G\widetilde G$ and $\theta_{\rm F}={\rm Arg}~{\rm det}(Y_uY_d)$ a function of the SM Yukawa couplings $Y_u\overline{q}\widetilde Hu+Y_d\overline{q}Hd$. Experimentally we find that~\cite{PDG}
\ba\label{robust}
\bar\theta<10^{-10}~~~~~~~~~~~~~~~~~~~~~~~~\theta_{\rm CKM}\sim1.
\ea
The first constraint follows from the current $90\%$ CL bound on the neutron EDM, $|d_n|\leq2.9\times10^{-26}~e$ cm, and the fact that the contribution of $\theta_{\rm CKM}$ to $d_n$ is very small~\cite{Ellis:1978hq}\cite{Khriplovich:1985jr}\cite{Czarnecki:1997bu}. Unfortunately the relation between $d_n, \bar\theta$ is not known to better than an order of magnitude (see e.g.~\cite{Vicari:2008jw}\cite{Engel:2013lsa} and references therein), so the constraint on $\bar\theta$ quoted in (\ref{robust}) --- obtained using naive dimensional analysis (NDA) --- suffers from a large uncertainty. 

The experimental inputs (\ref{robust}) are at the heart of the so called strong CP problem: how come $\bar\theta$ is so small despite the fact that CP is not a symmetry of the SM? Models that attempt to solve the strong CP problem via new symmetry principles can be broadly classified in two categories: 
\begin{itemize}
\item scenarios with a global $U(1)$ with a color anomaly~\cite{Peccei:1977hh}\cite{Weinberg:1977ma}\cite{Wilczek:1977pj}. If the symmetry is spontaneously broken at some high scale and explicitly broken dominantly by non-perturbative QCD effects, then the $\bar\theta$ angle dynamically relaxes to zero, and the theory predicts a QCD axion. %~\footnote{A composite axion also arises when promoting the Yukawas to dynamical fields. When restricted to the renormalizable part of the action, the axial symmetry is only violated by a term $\propto{\rm det}(Y)$ that implies $\theta=0$, see~\cite{Fong:2013sba}. However, higher order terms are un-suppressed, and spoil this conclusion.} 
The $\bar\theta$ angle becomes unphysical if $U(1)$ is not spontaneously broken.~\footnote{For earlier work on the controversial ``missing up-quark mass solution" see~\cite{mu}. This possibility is currently strongly disfavored by lattice data, see e.g.~\cite{Aoki:2013ldr}.} 
\item models with spontaneous CP~\cite{Nelson:1983zb}\cite{Barr:1984qx}\cite{Hiller:2001qg}\cite{Harnik:2004su}\cite{Cheung:2007bu}\cite{Antusch:2013rla} and/or P violation~\cite{Barr:1991qx}\cite{Babu:1989rb}\cite{Kuchimanchi:1995rp}\cite{Mohapatra:1995xd}\cite{Mohapatra:1997su}. Here $\bar\theta=0$ in the UV, and gets generated after spontaneous breaking.  
\end{itemize}
There are also examples in which both a mirror symmetry and an anomalous $U(1)$ are present, e.g.~\cite{Berezhiani:2000gh}\cite{Hook:2014cda}.

The attractive feature characterizing the QCD axion is that this approach does not restrict flavor nor CP violation in the UV: whatever source of CP violation is present will be washed out from the neutron EDM. The problem is that in order to evade current laboratory and astrophysics bounds, the axion decay constant has been pushed in a very uncomfortable regime in which the axion potential becomes enormously sensitive to possible $U(1)$-breaking trans-Planckian effects~\cite{AxionPlanck}. To some extent, the QCD axion has lost part of its original appeal, and has started to look more like a remarkable accident of the physics at very short distances. It is therefore useful to investigate the plausibility of alternative solutions.

CP is a gauge symmetry in several extra-dimensional extensions of the SM, including critical string theories~\cite{Choi:1992xp}\cite{Dine:1992ya}. This means that one can build scenarios based on spontaneous CP violation in which quantum gravity poses no threat to the basic symmetry principle. The question however is whether and how CP is broken after compactification down to our 4 space-time dimensions. It is conceivable that there exist a large number of realistic vacua in which CP breaking can be modeled via an effective 4D Lagrangian, but no definite conclusion can be drawn without a concrete model and a theory of moduli stabilization.

Nevertheless, the important point is that from an effective 4D theory perspective it is perfectly sensible, and also very well motivated, to assume that CP is a good symmetry of the UV, spontaneously broken at a scale parametrically low compared to the Planck scale. In this paper we will explore the viability of these models from a low energy perspective.

We assume CP is spontaneously broken in a secluded color-neutral sector and communicated to the SM via messenger fields, and ask: what constraints should the messenger dynamics satisfy in order to simultaneously account for a small neutron EDM and a large CKM phase? In section \ref{sec:general} we present sufficient conditions for this to happen. Our model-independent approach suggests novel solutions to the strong CP problem. In particular, we show that warped extra dimensions offer a variety of ways to implement our requirements, and analyze an explicit model in sec.~\ref{sec:5D}. Section~\ref{sec:pheno} highlights the main phenomenological signatures of spontaneous CP violation. A discussion of our results is given in section~\ref{sec:conclusions}.

\section{Model-independent analysis}
\label{sec:general}

\subsection{The CKM phase}

Let us denote by $\Sigma$ the CP-odd scalar(s), and assume CP violation is communicated to the SM quarks via an interaction with mediators characterized by a mass scale $m_*$. Collectively denoting by $\lambda$ the couplings involved, the size of CP violation in the visible sector is expected to be controlled by the following complex parameter $\xi\equiv\frac{\lambda\Sigma}{m_*}$. Indeed, this is what one finds in explicit scenarios of CP violation~\cite{Nelson:1983zb}\cite{Barr:1984qx}\cite{Hiller:2001qg}\cite{Harnik:2004su}\cite{Cheung:2007bu}. Crucially, the second experimental evidence in (\ref{robust}) {\emph{requires}} $\xi\sim1$, or more precisely
\ba\label{coincidence}
{\rm Re}(\xi),~{\rm Im}(\xi)\sim1.
\ea
Note that this implies the existence of a large mixing between the SM quarks and the messenger dynamics.

We review in Appendix~\ref{app:NB} and~\ref{sec:app} how (\ref{coincidence}) is realized in some of the existing models. In the following we will assume that $\xi\sim1$ has been arranged, and look at the first requirement in (\ref{robust}).

\subsection{A small $\bar\theta$}
\label{sec:spurion}

Finding reliable estimates for $\bar\theta$ is rather prohibitive, especially in view of two facts. First, the current bound is so stringent that even high-loop effects can spoil an otherwise brilliant solution. A direct estimate would therefore require a study of multi-loop diagrams, which is technically challenging. Second, we just saw that $\theta_{\rm CKM}\sim1$ implies a large quark-messenger mixing $\xi$ at the matching scale $m_*$, see (\ref{coincidence}), so no obvious expansion parameter is available.

\begin{figure}[t]
\begin{center}
\includegraphics[width=9cm]{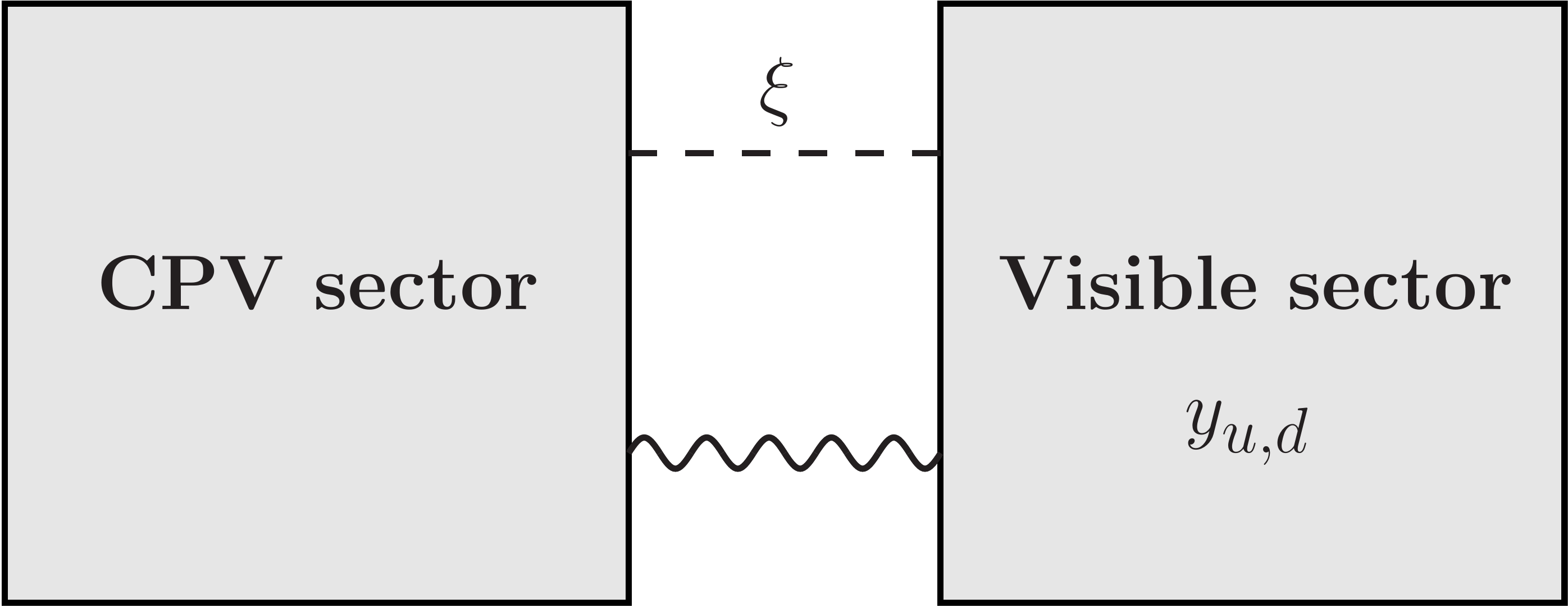}
\caption{Schematic picture of the setup: CP is spontaneously broken in a color-neutral CPV sector and communicated to the visible sector via a CP-odd and flavor-violating vev $\xi\sim1$. An example of minimal CPV sector is presented in Appendix~\ref{sec:conf}, while two models for the messenger dynamics are discussed in sec.~\ref{sec:5D} and Appendix~\ref{app:NB} (both belong to a more general class, see eq.~(\ref{CFT})). As argued at the end of this section, flavor-violation beyond MFV is severely constrained by~(\ref{robust}), whereas flavor-conserving interactions (wavy line) are not. 
}\label{fig}
\end{center}
\end{figure}

Here we take advantage of the selection rules associated to the {\emph{spurious}} charges of $\xi$ to identify {\emph{robust}} sufficient conditions for having a small $\bar\theta$ in a non-Supersymmetric framework (we will comment on SUSY in sec.~\ref{sec:conclusions}). We start with the most minimal class of models, in which (in the same field basis with $\theta_{\rm QCD}=0$) flavor is broken solely by 
\ba\label{yxi}
y_{u,d}~~~~~~~~~~~~~~~\xi.
\ea
Here $\xi$ is a {{complex}} matrix, whereas $y_{u,d}$ are {\emph{real}} proto-Yukawa couplings. {This is the minimal set of flavor-violating parameters, because $y_{u,d}$ are {\emph{necessary}} to generate the SM quark masses without fine-tuning,~\footnote{To avoid tuning, the quark masses must vanish with some parameter ${Y}$ that has $\beta_Y\propto{Y}$. It will soon be clear that $Y$ cannot be identified with (functions of) $\xi$. In this sense the choice~(\ref{yxi}) is the most minimal.} and must be real to avoid re-introducing a strong CP problem. This assumption is equivalent to having 3 flavor-violating matrices: the SM Yukawas $Y_{u,d}$ and $\xi'(\xi,Y)$.} Our setup is summarized in figure~\ref{fig}. The generalization to less minimal scenarios goes in the direction of increasing $\bar\theta$, as argued at the end of this section.

%%%%%%%%%%%%%%%%%%
%%%%%%%%%%%%%%%%%%
%%%%%%%%%%%%%%%%%%
%%%%%%%%%%%%%%%%%%

Now, $\xi$ will typically carry some charge under the SM flavor group. However, not all representations are equally viable. For example, if $\xi$ is a SM flavor singlet it is hard to imagine how to explain (\ref{robust}). Furthermore, a CP-odd vev in the bi-fundamental or a sextet $\xi$ would generically have an unsuppressed complex determinant, and would result in a large $\bar\theta$. The simplest non-trivial representations of the SM flavor group that are compatible with a $\theta_{\rm CKM}$ at leading order in $y$ are the triplet and octet of $SU(3)_Q$, with $Q=q,u,d$. We will therefore study scenarios with $\xi\sim{\bf 3}, {\bf 8}\in SU(3)_Q$.

We estimate $\bar\theta$ by matching the UV theory onto the SM at a scale $m_*\gg 100$ GeV. In this approach there are two types of contributions to $\bar\theta$. First, there are UV-sensitive local contributions, that appear as {\emph{polynomials}} of $y_{u,d}, \xi$ with real coefficients.~\footnote{In principle, non-perturbative (real) functions may also appear, but do not affect our conclusions.} Second, there are UV-finite corrections, which may contain more complicated functions (e.g. logs) of $y_{u,d}, \xi$. We postulate that the exotic fields have masses $\geq m_*$ that remain finite as $y,\xi\to0$. This assumption guarantees that all UV-finite effects decouple as $m_*\to\infty$ (and also that the CKM is unitary up to $m_{\rm SM}^2/m_*^2$ terms, see section~\ref{sec:pheno}).

%The present formalism provides an estimate for both UV-sensitive and finite contributions (see below).

The dominant flavor-singlet CP-odd combinations at leading order in the SM Yukawas are presented in table~\ref{table} for flavor-anarchic $\xi\sim{\bf 3}, {\bf 8}\in SU(3)_Q$. The factors of $4\pi^2$ in the table are estimated using NDA, but one should keep in mind that the actual numerical factors are model-dependent (see section~\ref{sec:estimate}).

%%%%%%%%%%%%%%%%%%%%%

\begin{table}[t]%[thh]
\begin{center}
\begin{tabular}{c|c|c} 
\rule{0pt}{1.2em}%
$\xi$ & Leading~CP-odd~structure & NDA estimate~of~$\bar\theta$ \\
%\hline
\hline
${\bf 3,8}\in SU(3)_q$ & ${\rm Im}~{\rm tr}\left\{f_1(y_u y_u^\dagger)f_2(y_d y_d^\dagger)-(1\leftrightarrow2)\right\}$
& $\lambda_C^2\frac{Y_t^2}{4\pi^2}\frac{Y_b^2}{4\pi^2}\sim5\times10^{-9}$\\
\hline
${\bf 3}\in SU(3)_u$ & ${\rm Im}~\left\{\xi^\dagger[y_u^\dagger y_u,y_u^\dagger y_dy_d^\dagger y_u]\xi\right\}$ & $\lambda_C^2\left(\frac{Y_t^2}{4\pi^2}\right)^2\frac{Y_b^2}{4\pi^2}\frac{Y_c}{Y_t}\sim3\times10^{-13}$\\
${\bf 8}\in SU(3)_u$ & ${\rm Im}~{\rm tr}\left\{(y_u^\dagger y_u)^2{\xi}[y_u^\dagger y_u,{\xi}]\xi\right\}$ & $\left(\frac{Y_t^2}{4\pi^2}\right)^2\frac{Y_c^2}{4\pi^2}\sim8\times10^{-11}$\\
\hline
${\bf 3,8}\in SU(3)_d$ & ${\rm Im}~{\rm tr}\left\{f_1(y_d^\dagger y_uy_u^\dagger y_d)f_2(y_d^\dagger y_d)-(1\leftrightarrow2)\right\}$ & $\lambda_C^2\frac{Y_t^2}{4\pi^2}\left(\frac{Y_b^2}{4\pi^2}\right)^2\frac{Y_s}{Y_b}\sim5\times10^{-16}$\\

\end{tabular}
\end{center}
\caption{\small Leading non-decoupling contributions to $\bar\theta$ in an expansion in $O(Y^2)$ for the models of fig.~\ref{fig}, with various representations of the CP-odd spurion $\xi$. The $f$s are real functions of $\xi$, with $f_{1}\neq f_2$. In the last column we estimated $\bar\theta$ using NDA, with the SM Yukawas $Y_{u,d}$ renormalized at $1$ TeV, $\lambda_C\simeq0.23$ the Cabibbo angle, and $\xi$ assumed to be flavor-anarchic with entries of order unity. As explained in the text, the leading CP-odd structures can be equivalently written by replacing $y\to Y$. 
\label{table}}
\end{table}
%%%%%%%%%%%%%%%%%%%%%%%

Consider for example $\xi$ in the adjoint of $SU(3)_q$. After integrating out the messenger dynamics of fig.~\ref{fig}, the SM Yukawas are
\ba
Y_{u,d}=f_{u,d}(\xi)y_{u,d}+O(y^3) 
\ea
for some function $f_{u,d}$ with real coefficients. A large CKM phase generically follows from ${\rm Arg}(\xi)\sim1$ and $f_u\neq f_d$.  (While intuitively clear, the reader can explicitly check this last statement using for example the results of~\cite{Rasin:1997pn}.) Since $\xi$ is hermitian, $f_{u,d}$ are as well, and we find that ${\rm det}Y$ is real at $O(y^2)$. At higher order, because of the identity ${\rm Arg}~{\rm det}(M)={\rm Im}~{\rm tr}\ln(M)$, $\theta_{\rm F}$ can be written as a trace of polynomials of $y_{u,d}, \xi$. Furthermore, we can use the Cayley-Hamilton identity to eliminate powers of matrices higher than 2.

In our field basis, under CP $y_{u,d}\to y_{u,d}^*=y_{u,d}$ are unchanged, and similarly for all other flavor-invariant couplings, whereas $\xi\to\xi^*=\xi^t\neq\xi$. There is no CP-odd polynomial  that involves only the CPV spurion, because any real function of a hermitian matrix has a real trace. To generate $\bar\theta$ we need insertions of the Yukawa matrices, which play the role of our small expansion parameters.

By counting the flavor indices it is easy to see that for $\xi\sim{\bf 8}\in SU(3)_q$ all flavor-singlets are traces times real determinants.~\footnote{Recall that pairs of Levi-Civita tensors with co-variant anti-covariant indices can be written as traces.} On the other hand, for $\xi\sim{\bf 3}\in SU(3)_q$ one can also find CP-odd expressions such as
\ba\label{up}
(y_uy_u^\dagger\xi)_i(y_dy_d^\dagger\xi)_j(y_uy_u^\dagger y_dy_d^\dagger\xi)_k\epsilon^{ijk},
\ea
where we used the fact that ${\rm det}(yy^\dagger)$ are real while $A_iA_j\epsilon^{ijk}=0$ for any vector $A$. To estimate its size (as well as the size of the invariants in the table) we observe that, because $y=f(\xi)Y+O(Y^3)$, replacing 
$$
y\to Y
$$ 
in the above expression results in sub-leading corrections. We can now perform flavor rotations to diagonalize $Y_u$ and put the down-type Yukawa in the form $VY_d$ with $Y_d$ diagonal and $V$ the CKM matrix. This has no effect on (\ref{up}), which is flavor singlet (though in general $\xi\to\xi'\neq\xi$). The numerical value of (\ref{up}) is smaller than that quoted in the first line of table~\ref{table}. Furthermore, we will see that in explicit scenarios, flavor triplet $\xi$s carry another spurious charge under a gauged $U(1)_\xi$, which actually forbids (\ref{up}) --- i.e. more powers of $y$ are necessary to build a singlet.

One can proceed analogously for the other representations shown in table~\ref{table}. In the case $\xi\in SU(3)_u$ we have ($f=c_1+c_2\xi\xi^\dagger$, with $c_{1,2}$ real functions of $|\xi|^2$)
\ba\label{Yu}
Y_u=y_uf(\xi)+O(y_{u,d}^3)~~~~~~~~Y_d=y_d+O(y_{u,d}^3).
\ea
Again, the contributions to $\bar\theta$ involving the Levi-Civita tensors are smaller than those of table~\ref{table} (i.e. a large number of $y$s are needed to build a singlet under $U(1)_\xi$ and the axial quark symmetry). Analogous considerations hold for $\xi\sim{\bf 8}\in SU(3)_u$ and $\xi\sim{\bf 3, 8}\in SU(3)_d$.

Looking at table~\ref{table} we conclude that theories with $\xi\sim{\bf 3}\in SU(3)_u$ or CP-odd spurions charged under $SU(3)_{d}$ are robustly consistent with data for generic $\xi\sim1$. In these scenarios the strong CP problem is simply addressed by realizing the framework depicted in fig.~\ref{fig}. A class of 4D theories satisfying these properties is discussed in Appendix~\ref{app:NB}. In section \ref{sec:5D} we will show how to construct alternative 5D realizations.

The viability of scenarios with $\xi\sim{\bf 8}\in SU(3)_u$ is somewhat model-dependent, whereas models with ${\xi}\in SU(3)_q$ can be made compatible with data by invoking an anti-correlation between $\xi, y_{u,d}$. For example, the powers of $Y_{t,b}$ in the first line of table~\ref{table} would be replaced by the Yukawas of the lighter generations if $(f_{1,2})_{3i}\ll1$ in the {\emph{very same basis}} with diagonal $Y$. Because this seems a rather fortunate coincidence, we will focus on $\xi\in SU(3)_{u,d}$ in the following.

\subsubsection{Generalization}

The above analysis is conservative, because it assumes the minimal ingredients (\ref{yxi}) needed to construct a realistic theory. We now want to study the impact of possible departures from this minimal framework.

First of all, models with multiple unsuppressed and uncorrelated CP-odd spurions with indices in the {\emph{same}} $SU(3)_Q$ are disfavored, because in that case $\bar\theta$ is typically renormalized at $O(y^2)$, e.g. $\bar\theta\propto{\rm tr}\left\{y^\dagger y[f_1,f_2]\right\}$ for $\xi\in SU(3)_{u,d}$. Similarly, models with $\xi$s in different $SU(3)$s are usually disfavored. The only exception are scenarios with {\emph{one}} $\xi_u\in SU(3)_u$ and {\emph{one}} $\xi_d\in SU(3)_d$, which are allowed. 

Furthermore, additional flavor-violating CP-even couplings must be small. A model-independent bound on the latter can be obtained as follows. Denote by $\Lambda_F$ the mass threshold at which new $O(1)$ sources of flavor violation arise. Then $\bar\theta$ will be generated by loops involving the SM, the messengers, and heavy particles of mass $\Lambda_F$. We can formally integrate out the CP-invariant flavor dynamics, obtaining (non-local) 4-fermion interactions such as $C_{ijkl}(\overline{q_i} Q_j)(\overline{Q_k} q_l)/\Lambda_F^2$. Closing a loop with the SM Yukawa, from the latter operators we estimate a 1-loop correction to $Y$ which translates into
\ba\label{PC}
\bar\theta=\frac{C}{4\pi^2}\frac{m_*^2}{\Lambda_F^2}\ln\frac{\Lambda_F}{m_*}.
\ea
For generic $C_{ijkl}\sim1$ one gets $C\sim Y_{Q_3}/Y_{Q_1}$, and (\ref{robust}) requires $m_*/\Lambda_F\lesssim5\times10^{-7}$. As a best case scenario we could imagine $\xi$ charged under $SU(3)_d$ and $C\sim Y_b^2\lambda_C^2$, in which case the bound (\ref{robust}) is satisfied for $m_*/\Lambda_F\lesssim10^{-2}$. No parametric suppression is available if $m_*\gtrsim\Lambda_F$. The lesson we learn is that a robust solution of the strong CP problem via spontaneous CP violation must satisfy $m_*/\Lambda_F\ll1$: flavor anarchy is severely constrained.

%%%%%%%%%%%%%%%%%%
%%%%%%%%%%%%%%%%%%
\begin{figure}[t]
\begin{center}
\includegraphics[width=4.5cm]{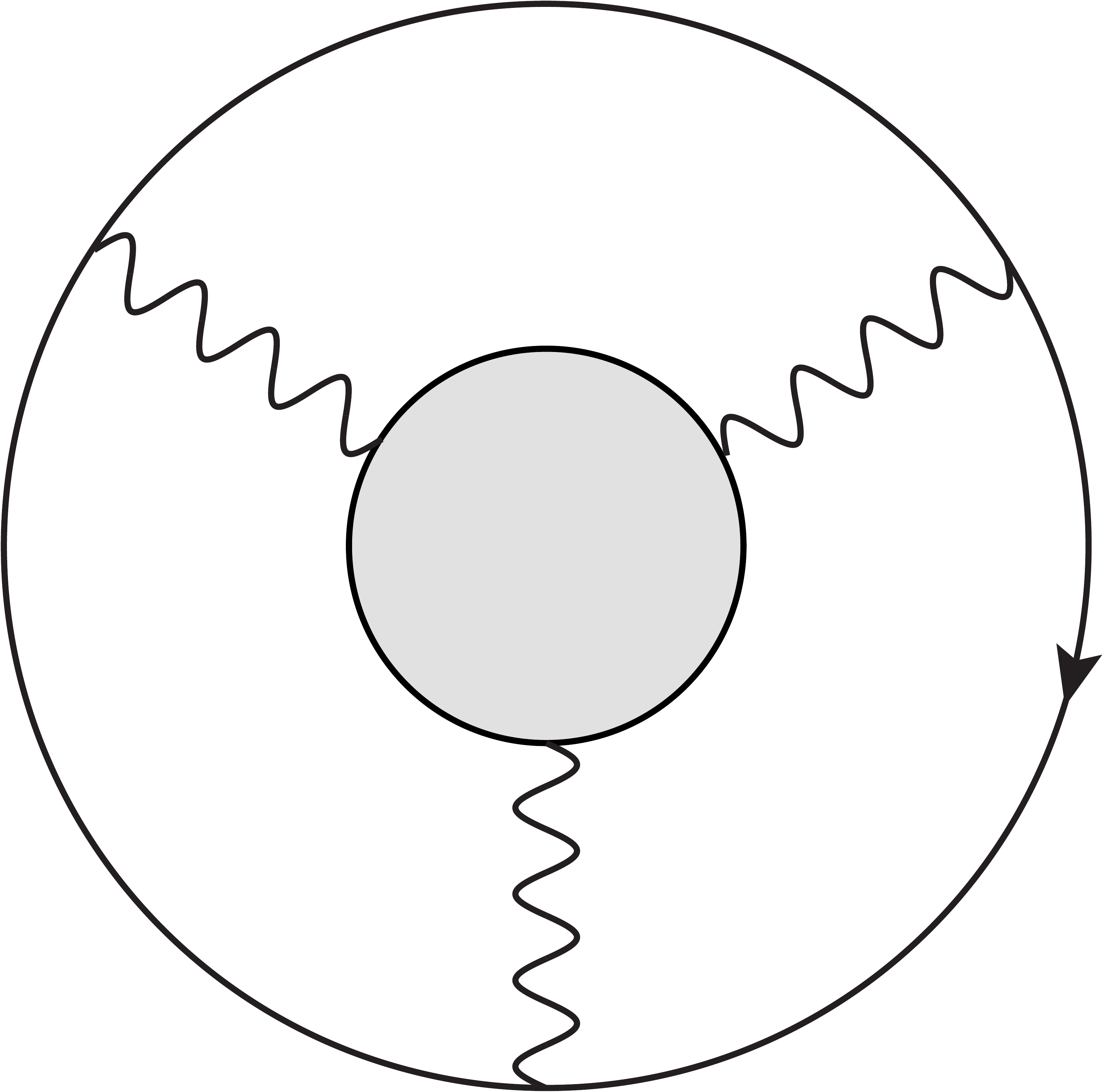}
\caption{This diagram illustrates how a CP-violating non-colored sector can contribute to $\bar\theta$ via gauge loops. The blob refers to loops of colorless particles, the wavy lines are gauge bosons, and the solid line is a colored fermion (quark or messenger). By cutting the solid line --- and attaching a Higgs if necessary ---, the diagram renormalizes the colored fermion mass. Alternatively, by attaching gluons to the solid line and an external $\xi$ to the blob, the graph becomes a direct contribution to $\bar\theta$ (see, e.g.,~\cite{Georgi:1980cn}).  
}\label{figLep}
\end{center}
\end{figure}
%%%%%%%%%%%%%%%%%%
%%%%%%%%%%%%%%%%%%

Finally, what about possible flavor-conserving interactions between quarks and the CPV sector? Let us switch off $\xi$, and assume that the colorless CPV dynamics couples to the SM quarks via the electro-weak interactions. Then, it is possible to see that the first Feynman diagrams that can induce a neutron EDM are formally encoded in fig.~\ref{figLep}, see the caption. Assuming for simplicity that all fields involved in the loop have comparable masses, we estimate 
\ba
\bar\theta\sim\left(\frac{g_w^2}{16\pi^2}\right)^3\bar\theta_{w}\sim10^{-10}~\left(\frac{\bar\theta_{w}}{5\times10^{-3}}\right),
\ea
where $\bar\theta_w$ is the largest CP-odd flavor-singlet in the CPV sector. Because in any perturbative theory $\bar\theta_{w}$ is {\emph{at most}} of 1-loop size, $\bar\theta_{w}\lesssim10^{-2}$, we conclude that the flavor-conserving interactions in fig.~\ref{fig} do not affect our conclusions.

\section{A realistic model on AdS$_5$}
\label{sec:5D}

The relation~(\ref{coincidence}) is basically the requirement that quarks mix at $O(1)$ with some exotic messenger sector. The problem of constructing theories of this type is therefore analogous to obtaining a large top mass in models with TeV scale compositeness. We illustrate this by constructing a realistic 5D model that simultaneously addresses the Planck-TeV hierarchy (as in~\cite{Randall:1999ee}\cite{Goldberger:1999uk}).

\subsection{Setup}

The basic ingredients needed to meet the criteria of fig~\ref{fig} are minimal flavor violation (MFV) deformed by the CP-odd (and flavor-violating) spurion $\xi$. There are several ways in which this can be realized on a warped extra dimension. MFV can be obtained as discussed using a CFT language in~\cite{Redi:2011zi}. We focus on a scenario with composite $u,d$, of which the 5D realization has not been explicitly presented yet. Models with composite $q$s or a fully composite SM are also possible, see e.g.~\cite{Rattazzi:2000hs}\cite{Cacciapaglia:2007fw}, but more constrained by electro-weak data.

Consider a slice of AdS$_5$ 
\ba
ds^2=a^2\left(\eta_{\mu\nu}dx^\mu dx^\nu-dz^2\right)~~~~~~~~~a=\frac{L}{z},
\ea
in the interval $z\in[z_{\rm UV}, z_{\rm IR}]$. Below we will see that the size of the extra dimension is stabilized such that $1/z_{\rm IR}\sim m_*\sim$ TeV with $1/z_{\rm UV}\sim\Lambda\sim10^{18}$ GeV. The bulk and the IR brane respect CP and the following {\emph{gauge}} symmetry 
\ba\label{global}
SU(3)_C\times SU(2)_w\times U(1)_Y\times G_F\times U(1)_\xi, 
\ea
where $G_F=SU(3)_u\times SU(3)_d$. It is straightforward to enlarge the gauge symmetry to account for a custodial $SU(2)$.

The UV brane violates $G_F$ at $O(1)$. On the other hand, CP$\times U(1)_\xi$ are spontaneously broken by a UV-localized scalar $\Sigma$ with charge $+1/2$ under $U(1)_\xi$, and a vev set by a scale $m_{\rm CP}\ll\Lambda$. This can be naturally achieved by promoting $\Sigma$ to a bulk scalar of another AdS$_5$ throat, or more simply by identifying $\Sigma$ with a fermion bilinear, as shown in Appendix~\ref{sec:conf}. The additional requirement $m_{\rm CP}\gg T_{\rm RH}\gg 1/z_{\rm IR}$, with $T_{\rm RH}$ the re-heating temperature, guarantees that a potential domain wall problem associated to the spontaneous breaking of CP is evaded. This requirement motivates our choice of localizing $\Sigma$ on the UV, rather than in the bulk, though we emphasize that the strong CP problem may still be solved with a bulk $\Sigma$.

The CP-odd scalar has very suppressed couplings to the SM fermions, that are $U(1)_\xi$ neutral (see below). CP violation shines through the bulk thanks to the UV-localized coupling 
\ba\label{O}
\left.\lambda\Sigma\phi\right|_{z_{\rm UV}},
\ea
with $\phi$ a bulk scalar in the fundamental representation of $SU(3)_u$ and transforming with charge $-1/2$ under $U(1)_\xi$. We will show that this will lead to $\xi\propto\lambda\langle\Sigma\rangle\sim{\bf 3}\in SU(3)_u$.

\subsection{Minimal flavor and CP violation}

There are 4 types of bulk fermions:~\footnote{With this field content, there are $G_F^3, G_F^2\times U(1)_Y$ anomalies, which we assume are cancelled by UV-localized fermions. These acquire masses of order $\Lambda$ after $G_F$ breaking, and have no phenomenological impact.} $\psi_{u,d}$, that have the same SM charges of the right-handed quarks $u,d$, and two weak doublets $\psi_{q_u,q_d}$ with the same SM charges as $q$. They transform under $G_F$ as $\psi_{u,q_u}\sim{\bf 3}\in SU(3)_u$, $\psi_{d,q_d}\sim{\bf 3}\in SU(3)_d$. All fermions are $U(1)_\xi$-singlets. 

The boundary conditions are
\ba\label{BC}
z_{\rm IR}~\left\{ \begin{array}{c}  
\left.\psi_{u,d}\right|_L=0 \\
\left.\psi_{q_u,q_d}\right|_R=0
\end{array}\right.~~~~~~~~~~~~~~
z_{\rm UV}~\left\{ \begin{array}{c}  
\left.\psi_{u,d}\right|_L=0 \\
\left.\psi_{q_u}\right|_L=F\left.\psi_{q_d}\right|_L
\end{array}\right.
\ea
where $F$ is a 3 by 3 $G_F$-violating matrix, real by CP. The SM $u,d$ arise as the zero modes of $\psi_{u,d}$, whereas a single left-handed combination of the doublets survives (\ref{BC}) and will be identified with $q$. 

The ($G_F$-symmetric) bulk fermion masses are chosen so that the zero-modes of $\psi_{u,d}$ peak in the IR, whereas $\psi_{q_u,q_d}$ are localized towards the UV. In terms of the bulk mass parameters $c=m_5L$ this reads $c_{u,d}<1/2$, $c_{q_u,q_d}>1/2$. 

Consistently with our assumptions we add $G_F$-violating UV-localized kinetic terms for the fermions
\ba
{\cal L}_{\rm kin}=\sum_{Q=u,d,q_u,q_d}\int dz~a^4(z)~\overline{\psi_Q} K_Q i\slashed{D}\psi_Q~\delta(z-z_{\rm UV}),
\ea
with $K_Q$ symmetric (real) matrices, as well as a UV mass for the $G_F$ gauge field, which ensures that the low energy theory has no exotic massless vectors. 

The kinetic terms affect the normalization of the 4D modes. However, because the heavy modes are localized towards the IR brane, the effect of the UV kinetic terms is suppressed by powers of $z_{\rm UV}/z_{\rm IR}$ and is numerically negligible. Similar considerations hold for the zero-modes of $\psi_{u,d}$. It follows that $G_F$-violation is entirely encoded in the UV kinetic terms and UV boundary conditions of the zero-modes of $\psi_{q_u,q_d}$.

The wave-function of the left-handed zero-modes read 
\ba\label{NF}
\psi^0_{q_u}=\frac{1}{\sqrt{L}}\left(\frac{z}{z_{\rm UV}}\right)^{2-c_{q_u}}FN~q(x)~~~~~~~~~~\psi^0_{q_d}=\frac{1}{\sqrt{L}}\left(\frac{z}{z_{\rm UV}}\right)^{2-c_{q_d}}N~q(x), 
\ea
where the normalization $N$ is a matrix in flavor space that satisfies 
\ba
1&=&N^\dagger\left[{\cal K}_{q_d}+F
^\dagger {\cal K}_{q_u}F\right]N,~~~~~~
{\cal K}_Q\equiv a^4(z_{\rm UV})\left[\frac{z_{\rm UV}}{L}\frac{\left(\frac{z_{\rm IR}}{z_{\rm UV}}\right)^{1-2c_Q}-1}{1-2c_Q}+K_Q\right].
\ea
This guarantees that $q$ has a canonical kinetic term in the effective 4D theory. (We neglected IR kinetic terms, that are flavor-diagonal and irrelevant to our discussion.)

In the effective field theory, flavor-violation is controlled by the two matrices $FN, N$, which may be interpreted as spurions transforming respectively as a $({\bf \overline{3},3,1})$ and $({\bf \overline{3},1,3})$ under $SU(3)_q\times SU(3)_u\times SU(3)_d$. The resulting model satisfies minimal flavor (and CP) violation. In particular, the SM Yukawa couplings arise from the IR-localized operators $Y_{u*}L\overline{\psi_{q_u}}\widetilde H\psi_u+Y_{d*}L\overline{\psi_{q_d}}H\psi_d$, with $Y_{u*,d*}$ real numbers, and after KK reduction the proto-SM Yukawa matrices are
\ba
y_u&=&Y_{u*}N^\dagger F^\dagger\sqrt{1-2c_u}\left(\frac{L}{z_{\rm UV}}\right)^{3/2}\left(\frac{z_{\rm UV}}{z_{\rm IR}}\right)^{c_{q_u}-1/2}\\\no
y_d&=&Y_{d*}N^\dagger\sqrt{1-2c_d}\left(\frac{L}{z_{\rm UV}}\right)^{3/2}\left(\frac{z_{\rm UV}}{z_{\rm IR}}\right)^{c_{q_d}-1/2},
\ea
up to hierarchy-suppressed corrections. Here for simplicity we assumed the Higgs boson is IR-localized, but one can consider a bulk scalar or a Nambu-Goldstone Higgs. Furthermore, note that because of the suppression $({z_{\rm UV}}/{z_{\rm IR}})^{c-1/2}$ the quark masses can be obtained with ${K}_Q=O(1)$ (though no explanation of the hierarchy is offered).

\subsection{$\theta_{\rm CKM}$ and moduli stabilization}
\label{sec:stab}

We anticipated that CP is violated on the UV and shined through the bulk using a bulk scalar $\phi\sim{\bf 3}\in SU(3)_u$. The Lagrangian of $\phi$ reads
\ba\label{phiaction}
{\cal L}_\phi&=&\int dz~a^3\left[\eta^{\mu\nu}D_\mu\phi^\dagger D_\nu\phi-D_5\phi^\dagger D_5\phi-{a^2m_\phi^2}\phi^\dagger\phi+\cdots\right]\\\no
&+&a^4\left.\left[-\frac{T_{\rm UV}}{L^4}+\left(\frac{J^\dagger}{L^{5/2}}\phi+{\rm hc}\right)+\cdots\right]\right|_{z_{\rm UV}}\\\no
&+&a^4\left.\left[-\frac{T_{\rm IR}}{L^4}-\frac{m_{\rm IR}}{L}\phi^\dagger\phi+\cdots\right]\right|_{z_{\rm IR}},
\ea
where $\cdots$ refer to higher order couplings. The non-linear bulk couplings of $\phi$ are assumed to be somewhat suppressed, while the boundary terms have generic coefficients (this is possible by locality of the 5D theory). 

Because $m_\phi^2>-4$, higher-dimensional operators on the UV brane are naturally suppressed by powers of the hierarchy and can be neglected. The dominant UV interaction is the source term, that encodes the coupling to the UV-localized CP-violating sector (see~(\ref{O}))
$$
J^\dagger=L^{5/2}\lambda\langle\Sigma\rangle.
$$
As shown in Appendix~\ref{sec:conf}, the dynamics of $\Sigma$ is completely irrelevant to us, only the vacuum $\langle\Sigma\rangle$ has a phenomenological impact. In what follows we will treat $J$ as a dimensionless {\emph{complex}} constant with $|J|\ll1$.

The source induces a complex vacuum expectation value for $\phi$, which eventually feeds into $\theta_{\rm CKM}, \bar\theta$. Imposing Neumann conditions on the IR we find, up to $O(1/M_5^3)$ with $M_5$ the 5D Planck scale,
\ba\label{sol}
\phi(z)&=&\frac{J}{\epsilon L^{3/2}}\left(\frac{z}{z_{\rm UV}}\right)^{-\epsilon}\frac{1-\eta\left(\frac{z}{z_{\rm IR}}\right)^{4+2\epsilon}}{1+\frac{4+\epsilon}{\epsilon}\eta\left(\frac{z_{\rm UV}}{z_{\rm IR}}\right)^{4+2\epsilon}}
~~~~~~~~\eta=\frac{m_{\rm IR}-\epsilon}{4+m_{\rm IR}+\epsilon},
\ea
where $\epsilon=-2+\sqrt{4+m_\phi^2L^2}$.

To communicate CP violation to the SM we need to couple $\phi$ to the SM quarks. A simple option is to add a fermion $\Psi$ in an appropriate representation of the gauge group to allow a trilinear coupling among $\phi,\Psi$ and $\psi_u$. The boundary conditions for $\Psi$ are chosen so that there are no zero modes. Integrating out $\Psi$ one gets a correction to the kinetic term of $\psi_u$ of order $\xi\xi^\dagger$, where $\xi\sim{\lambda_*\phi(z_{\rm IR})L^{3/2}}/{m_\Psi{z_{\rm IR}}}$, with $\lambda_*$ the dimensionless trilinear coupling. Since $m_\Psi\sim\pi/z_{\rm IR}$, we see that $\theta_{\rm CKM}\sim1$ is obtained for $\phi(z_{\rm IR})L^{3/2}=O(1)$, at least as long as $\lambda_*$ is not too small (for large $\lambda_*$ the CKM phase is effectively generated by higher dimensional operators, as discussed below). This mechanism is analogous to that of~\cite{Nelson:1983zb}\cite{Barr:1984qx} (see Appendix~\ref{app:NB}), although here (\ref{tune}) does not require additional model-building.

One can envision other couplings between $\phi$ and the SM quarks that ensure $\theta_{\rm CKM}\sim1$ equally well, and this shows that in general these models are not in the Nelson-Barr class. A large CKM phase may be parametrized in a model-independent way by IR-localized higher dimensional operators such as %(0907.0474)
\ba\label{HDO}
c_{{\rm IR}_{u,d}}\frac{16\pi^2}{\Lambda_5^4}\overline{\psi_{q_u,u}}\phi \phi^\dagger i\slashed{D} \psi_{q_u,u},~~~~~~~~~~~~~
c_{Y_u}\frac{(16\pi^2)^{3/2}}{\Lambda_5^4}\overline{\psi_{q_u}} \phi \phi^\dagger\widetilde H\psi_u,
\ea
where according to NDA $c_{{\rm IR}_{u,d}, Y_u}<1$. The 5D cutoff can be written as $\Lambda_5=N_{\rm KK}\frac{\pi}{L}$, where $N_{\rm KK}$ is a measure of the number of weakly coupled KK modes that can be described by our effective theory. Inserting the vev of $\phi$ one finds a correction to the Yukawa coupling of the form anticipated in (\ref{Yu}) with $\xi\xi^\dagger\propto\phi(z_{\rm IR})\phi^\dagger(z_{\rm IR})/{\Lambda_5^4L}$. We now need $\phi(z_{\rm IR}) L^{3/2}\sim5(N_{\rm KK}/5)^{2}$ to generate a large CKM phase.

We conclude that $\theta_{\rm CKM}\sim1$ generically follows provided $\phi(z_{\rm IR})L^{3/2}=O(1)$, which via (\ref{sol}) represents a non-trivial relation between $|J|\ll1$ and $z_{\rm UV}/z_{\rm IR}\ll1$. A nice feature of these 5D models is that this relation can be naturally achieved if $\phi$ is responsible for stabilizing the extra dimension. Following~\cite{Goldberger:1999uk}\cite{Rattazzi:2000hs} we determine the size of the extra dimension by minimizing the potential ${\cal V}_{\rm eff}$ of $\mu=1/z_{\rm IR}$. Inserting (\ref{sol}) into (\ref{phiaction}) we get:
\ba
{\cal V}_{\rm eff}(\mu)&\equiv&-\left.{\cal L}_\phi\right|_{\rm on-shell}%=\mu^4 T_{\rm IR}+\mu_0^4\left(T_{\rm UV}-J^\dagger \phi(z_{\rm UV})L^{3/2}\right)
\\\no
&=&\mu^4\left[T_{\rm IR}+\frac{(4+2\epsilon)}{\epsilon^2}\eta J^\dagger J\left(\frac{\mu}{\mu_0}\right)^{2\epsilon}\right]+\mu_0^4\left[T_{\rm UV}-\frac{J^\dagger J}{\epsilon}\right]+O\left(\left({\mu}/{\mu_0}\right)^{8+4\epsilon}\right),
\ea
where $\mu_0=1/z_{\rm UV}$. In our model the operator dual to $\phi$ is relevant (i.e. $\epsilon<0$) and $|\epsilon|<1$, so a stable vacuum requires $T_{\rm IR}>0$ and $\eta<0$ (which is equivalent to $-4-\epsilon<m_{\rm IR}<\epsilon$). The minimum, up to $O\left(\left({\mu}/{\mu_0}\right)^{4+2\epsilon}\right)$, is determined by:
\ba\label{scale5D}
\frac{J^\dagger J}{\epsilon^2}\left(\frac{\mu}{\mu_0}\right)^{2\epsilon}=\frac{\phi^2(z_{\rm IR})L^3}{(1-\eta)^2}&=&-\frac{4}{(4+2\epsilon)^2}\frac{T_{\rm IR}}{\eta}.
\ea
As anticipated, we see that $\theta_{\rm CKM}=O(1)$ is obtained for $T_{\rm IR}\sim1$.

What we have just reviewed is the familiar Goldberg-Wise mechanism~\cite{Goldberger:1999uk}\cite{Rattazzi:2000hs} with a \emph{naturally small} (by the $U(1)_\xi$) UV source, and a relatively large $|\epsilon|$, say $\sim0.3-0.4$. The effective potential ${\cal V}_{\rm eff}$ was calculated ignoring gravity fluctuations, which are down by powers of $\phi^2/M_5^3\lesssim1/(M_5L)^3$. We explicitly checked that our result $\xi\sim1$ is not spoiled for $\phi^2\sim M_5^3$, nor for generic IR-localized $\phi$ potentials. The reason may be ultimately understood using the CFT language of the next subsection.~\footnote{As usual, the scalar spectrum of the model contains a tower of heavy modes of mass $\sim\pi/z_{\rm IR}$ and a light radion. In the limit of small back-reaction the latter can be efficiently interpolated by $\mu$ (see for instance~\cite{Vecchi:2010gj}). Defining the 5D Einstein-Hilbert action by $\int \sqrt{|g|}[-M_5^3R]$, and identifying $z_{\rm UV}=L$, we find that the kinetically normalized radion is $\sqrt{6(M_5L)^3}\mu$, and $m_{\rm rad}^2=\frac{4}{3}\frac{T_{\rm IR}}{(M_5L)^3}\frac{|\epsilon|}{z^2_{\rm IR}}$.}

\subsection{CFT interpretation, and estimate of $\bar\theta$}
\label{sec:estimate}

To show how robust the solution of the strong CP problem is, it is instructive to estimate $\bar\theta$ using a general CFT language.

The models of section~\ref{sec:5D} and Appendix~\ref{app:NB} belong to a larger class of scenarios realizing the framework of figure~\ref{fig}, that are described by a (large $N$) CFT with global symmetry $\supset SU(3)_{u,d}$, and deformed by the couplings:
\ba\label{CFT}
\delta{\cal L}_{\rm CFT}=y_uO_u+y_dO_d+\lambda\Sigma O.
\ea
In our model the global symmetry is (\ref{global}), $O_{u,d}=\overline{q}{\cal O}_{u,d}$, with ${\cal O}_{u,d}$ fermionic operators of the CFT with scaling dimension $2+c_{q_u,q_d}>5/2$, whereas $O$ is a scalar of dimension $d_O=4+\epsilon<4$ dual to $\phi$. Higher dimensional operators, and loops of the CP-violating dynamics, can be rendered completely negligible under very generic conditions on $m_{\rm CP}/\Lambda, d_O, d_\Sigma$, with $d_{\Sigma}$ the scaling dimension of $\Sigma$ (see Appendix~\ref{sec:conf}).

To generate a large CKM phase, $O$ must be the {\emph{most relevant}} deformation of the CFT, $d_O<4$. After symmetry breaking, the CP-violating coupling $\lambda\langle\Sigma\rangle$ grows towards the IR until the scale $m_*$ defined by
\ba\label{scale}
\xi\equiv J\left(\frac{m_*}{\Lambda}\right)^{d_O-4}\sim1,
\ea
with $J\equiv\lambda\left(m_{\rm CP}/{\Lambda}\right)^{d_\Sigma}$. This is the CFT dual of (\ref{scale5D}), with $m_*\sim1/z_{\rm IR}$. Note that $\xi$ has the same spurious charges as $O$ under $G_F$ rotations. Below $m_*$, CP violation is controlled by $\xi$, and a large CKM phase is expected provided the CFT is sufficiently generic.

The setup of figure~\ref{fig} is realized, and we expect $\bar\theta$ can be under control choosing appropriate charges for $O$ (in the present 5D model ${\bf 3}\in SU(3)_u$). Importantly, the expansion parameter here is truly the mixing $y^2/g^2_*$ between $q$ and the heavy resonances (KK modes) --- with $g_*$ the typical coupling among resonances of the CFT --- and can be much larger than $y^2/4\pi^2$, which we would find in a renormalizable 4D model.

In our model $\bar\theta$ cannot arise at tree-level. Tree-level corrections to ${\rm det}(Y)$ are written as traces of polynomials of $y_uD_uy_u^\dagger, y_dD_dy_d^\dagger$, where $D_u(\xi\xi^\dagger)$ and $D_d\propto1$ are the propagators of ${\cal O}_u$ and ${\cal O}_d$, but these do not renormalize $\bar\theta$ because the polynomials are symmetric under the exchange of the building blocks. To get $\bar\theta$ one needs insertions of 4-point functions involving ${\cal O}_{u,d}$, which suppress the amplitude by at least a factor of $g_*^2/4\pi^2$. Actually, neglecting higher-dimensional operators, in our 5D model the first contribution to $\bar\theta$ arises at 2-loop order via $\Psi\phi\psi_u$ and the Yukawa coupling, and is therefore further suppressed by a factor $g_*^2/4\pi^2$ compared to the generic CFT case. This gives $\bar\theta\sim10^{-11}(1/g_*)^2$, implying that these theories are capable of solving the strong CP problem even in the large $N$ regime $g_*\lesssim1$. Consistently, a similar estimate is obtained when turning on higher-dimensional operators. Diagrammatically, the loop effects can be seen as corrections to $\theta_{\rm F}$ or direct contributions to $\xi\frac{d\bar\theta}{d\xi}G\widetilde G$~\cite{Georgi:1980cn}.

\section{Phenomenology}
\label{sec:pheno}

In this section we discuss signatures that {\emph{decouple}} with the new physics scale $m_*$.

\subsection{Electro-weak observables}

Models with spontaneous CP violation require a large mixing between right-handed quarks and a messenger sector, see (\ref{coincidence}). In the 4D models of Appendix~\ref{app:NB} this is simply $\sim\xi m_*\overline{u}\Psi$ (or $\xi m_*\overline{d}\Psi$). Relevant contributions to the electro-weak observables arise at 1-loop, and are within current bounds for $m_*$ above several hundred GeV (see for instance~\cite{Aguilar-Saavedra:2013qpa}).

In the 5D models, the entire bulk plays the role of the messenger dynamics. As is well know, warped 5D scenarios are subject to severe bounds from electro-weak data. For example, writing the $S$ parameter as $\Delta S=8\pi v^2/m_*^2$ and imposing $\Delta S\lesssim0.2$~\cite{Baak:2012kk}, we find $m_*\gtrsim2.8$ TeV. Stronger constraints arise from $Z^0$ pole observables on models with ``composite doublets", $\xi\in SU(3)_q$~\cite{Redi:2011zi}.

\subsection{Flavor violation}

Flavor violation is a generic implication of our scenarios, with the dominant effects beyond MFV controlled by $\xi$. In the 4D models discussed in Appendix~\ref{app:NB}, $\Delta F=2$ transitions are controlled by a loop-induced $(\overline{q}\gamma^\mu YY^\dagger q)^2$, and current Kaon data are satisfied for $m_*\gtrsim1$ TeV~\cite{Bona:2007vi}\cite{Isidori:2010kg}, even in the worst case scenario $\xi\sim{\bf 3}\in SU(3)_u$. $B\to X_{s,d}$ transitions are also relevant, and controlled by 
\ba
\overline{q_d}[V^\dagger Y_ufY_u^\dagger VY_d]d.
\ea
Using standard results (see e.g.~\cite{Grinstein:1990tj}), we find that the main effect is described by the operator $Q_{7\gamma}=\frac{G_F}{\sqrt{2}}V_{ts}^*V_{tb}\frac{e}{4\pi^2}m_b\overline{s_L}\gamma^{\mu\nu}b_RF_{\mu\nu}$, with a coefficient $\delta C_{7\gamma}(m_*)\sim{m_t^2}/(3{m_*^2V_{ts}^*V_{tb}})$. Conservatively requiring $|\delta C_{7\gamma}(3~{\rm GeV})|\lesssim0.1$ sets again a bound on the messenger scale of order $m_*\gtrsim1$ TeV.

We now discuss flavor-violation in 5D models with $\xi\in SU(3)_u$, since scenarios with $\xi$ charged under $SU(3)_{d}$ are generically subject to more severe constraints from Kaon physics. Flavor-violating amplitudes in 5D models with $\xi\sim{\bf 3, 8}\in SU(3)_u$ can be written as polynomials with real coefficients of:
\ba
Y_d,~~~~~~~~~V^\dagger Y_u,~~~~~~~\xi
\ea
where $Y_{u,d}$ are diagonal SM Yukawas (with negligible $\bar\theta$) and $V$ the CKM matrix. The analysis is a slight generalization of that in~\cite{D'Ambrosio:2002ex}, where $\xi=0$. For example, all bounds on $\Delta F=1,2$ processes involving $q,d$ derived there immediately apply here as well. The exception are operators involving up type quarks. The most severely constrained is by far the $\Delta F=2$ interaction:
\ba\label{uc}
\frac{g_*^2}{m_*^2}(\overline{u_R}\gamma^\mu c_R)^2.
\ea
These have unsuppressed Wilson coefficients in a generic dynamics with anarchic $\xi$. A conservative bound from $D^0-\overline{D^0}$ mixing gives $m_*\gtrsim g_*\times10^3~{\rm TeV}$~\cite{Bona:2007vi}\cite{Isidori:2010kg}.

\begin{figure}[t]
\begin{center}
\includegraphics[width=8cm]{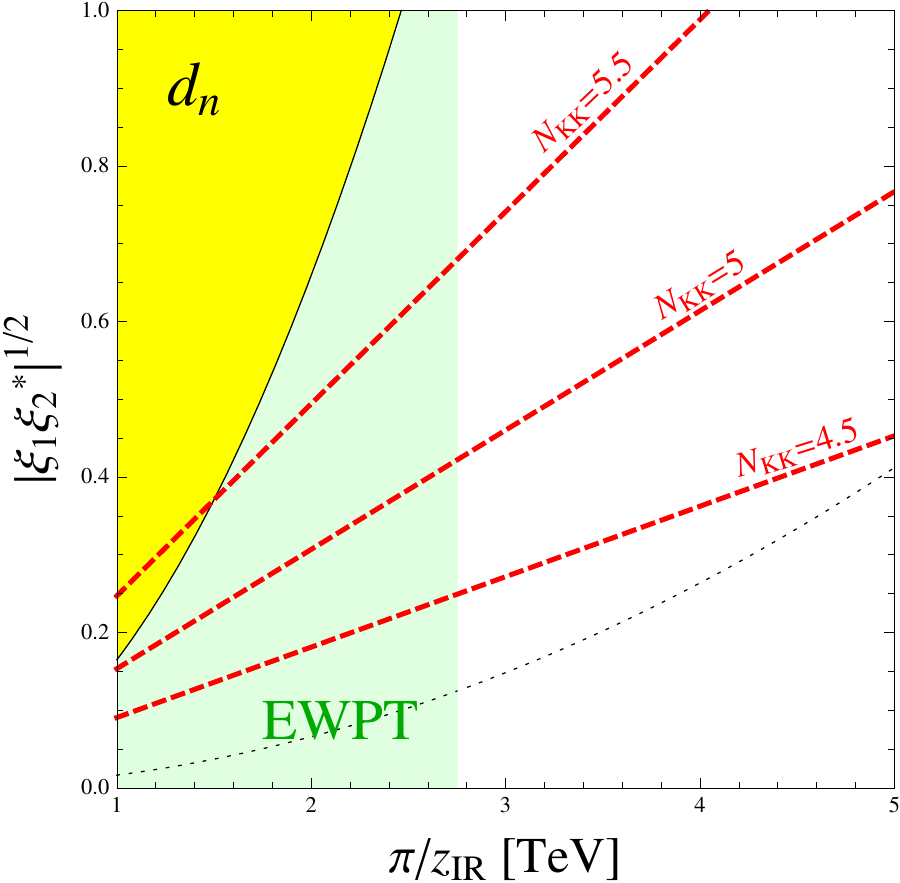}
\caption{Constraints from $d_n$, electro-weak, and flavor data in 5D models with $\xi\sim{\bf 3}\in SU(3)_u$. The contribution of the dim-6 operators to $d_n$ is dominated by the {\emph{second}} term in (\ref{dnq}). We took $|b_u(f_3)_{31}|=|\xi_1|$ and $d_n={\rm Im}(d_u)_{11}$, with $\xi_i$ defined in the basis with diagonal $Y_u$. The yellow area is excluded by current bounds, whereas the dotted black line shows the limit for a bound on $d_n$ $10$ times stronger. An indicative measure of the electro-weak constraints on $\pi/z_{\rm IR}$ is shown by the light green area. A conservative bound from flavor physics can be obtained assuming that the IR-localized operator $(\overline{\psi_u}\gamma^\mu\phi\phi^\dagger\psi_u)^2$ is generated at the 5D cutoff with a coefficient of order $(16\pi^2)^3/\Lambda_5^{10}$, which is the maximal value allowed by calculability. The parameter space excluded by imposing the constraint from $D^0-\overline{D^0}$ is above the dashed red lines, obtained with $\Lambda_5L=N_{\rm KK}\pi$ and $\phi^2(z_{\rm IR})L^3=1$.}\label{figB}
\end{center}
\end{figure}

A quantitative estimate of the bound can be obtained for the scenario of section~\ref{sec:5D}. In that case (\ref{uc}) receives corrections at tree-level by the $G_F$ gauge bosons, that have non-universal masses $\propto\xi$. However, these can be suppressed by simply assuming a small 5D gauge coupling. A reasonable expectation is that (\ref{uc}) is dominantly generated by physics above the cutoff $\Lambda_5$. The leading IR-localized operators (recall that $\phi$ is suppressed in the bulk) for $\xi\sim{\bf 3}\in SU(3)_u$ schematically look like $(\overline{\psi_u}\gamma^\mu\phi\phi^\dagger\psi_u)^2$. Because of the large engineering dimension, these can easily be under control if $\phi L^{3/2}\sim1$ and KK masses not far above the TeV, see fig.~\ref{figB}. Yet, for $\phi L^{3/2}\sim1$ also the higher-dimensional interactions (\ref{HDO}) are suppressed, and $\theta_{\rm CKM}$ must be generated by the tree-level exchange of a messenger $\Psi$ as explained above (\ref{HDO}).

%%%%%%%%%%%%%%%%%%
%%%%%%%%%%%%%%%%%%
%%%%%%%%%%%%%%%%%%
%%%%%%%%%%%%%%%%%%

Alternatively, we may relax the assumption of flavor anarchy for $\xi$. After all, flavor violation is controlled by unknown UV physics which must generate hierarchical SM Yukawas. It does not seem un-reasonable that $\xi$ is also hierarchical in the same field basis in which the $Y_{u,d}$ are. For example, a structure like $(y_{u,d})_{ij}\sim\epsilon_i^q\epsilon_j^{u,d}$ and $\xi_{i}\sim\epsilon^u_i$, with hierarchical $\epsilon$s, can still generate $\theta_{\rm CKM}\sim1$ but dramatically relaxes the bounds on (\ref{uc}). On the other hand, the estimates in table \ref{table} depend on the mixing between the second and third generations, and are therefore unaffected. In this class of models $t\to c$ transitions are a key signature, and $\theta_{\rm CKM}$ may well be generated by the higher dimensional operators in (\ref{HDO}).

\subsection{Neutron EDM (dimension 6)}

In addition to $\bar\theta$, there are higher dimensional operators that contribute to the neutron EDM, and can easily dominate.~\footnote{Recall that, assuming $\bar\theta=0$, the contribution from pure SM physics is expected to be of order $|d_n|/e\sim10^{-32}$ cm (see e.g.~\cite{Seng:2014lea} and references therein).} At dimension six the most relevant are $GGG$, $H^\dagger HG\widetilde G$, quark dipoles, and 4-fermion interactions. The first class of operators is flavor-conserving, very much like $G\widetilde G$, so its effect is suppressed by $\sim\Lambda_{\rm QCD}^2/m_*^2$ compared to $\bar\theta$ and can be neglected. 

The bounds on the dipoles are model-dependent. We start with $\xi\sim{\bf 3, 8}\in SU(3)_u$ and consider $d_u\overline{q}\widetilde H\sigma^{\mu\nu}u F_{\mu\nu}$. In realizations of the setup of figure~\ref{fig}, the neutron EDM is of order (the leading $O(Y)$ terms do not contribute, since the imaginary part of $(y_uf)_{ii}$ vanishes because the $f$s are hermitian)
\ba\label{dnq}
{{\rm Im}(d_u)_{11}}&=&\frac{{e}}{4\pi^2}\frac{v}{m_*^2}{\rm Im}\left[a_uY_uf_1Y_u^\dagger Y_u f_2+b_uVY_dY_d^\dagger V^\dagger Y_u f_3+O(Y^5)\right]_{11}\\\no
&=&a_ue\frac{m_{u}}{m^2_*}\frac{Y_{u_k}^2}{4\pi^2}{\rm Im}[(f_1)_{1k}(f^*_2)_{1k}]+b_ue\frac{m_{u_i}}{m^2_*}\frac{Y_{d_k}^2}{4\pi^2}{\rm Im}[V_{1k}V^*_{ik}(f_3)_{i1}]+O(Y^5),
\ea
where $a_u, b_u$ are model-dependent real numbers $\sim1$ and the $f$s polynomials of $\xi$ with real coefficients. For flavor-anarchic $f$s of order unity, the $90\%$ CL bound $|d_n|/e<2.9\times10^{-26}$ cm~\cite{PDG} becomes $m_*\gtrsim\sqrt{|a_u|}3.7, \sqrt{|b_u|}2.5$ TeV. Yet, in the 4D models of Appendix~\ref{app:NB} we find $f_1\propto\xi\xi^\dagger$, $f_{2,3}\propto1_{3\times3}$, and  the bound is negligible. The dipole for the down quark leads to analogous results.

Similar considerations hold for $\xi\sim{\bf 3, 8}\in SU(3)_d$ if we replace $Y_u\leftrightarrow VY_d$ in (\ref{dnq}), and we find $m_*\gtrsim\sqrt{|a_d|}0.09, \sqrt{|b_d|}20$ TeV for generic $f$s. In the case $\xi\sim{\bf 3, 8}\in SU(3)_q$ all corrections are proportional to the light mass, and $m_*$ above the TeV would suffice. 
 
The CP-odd 4-fermion interactions contributing to $d_n$ are of the form $\overline{q}u \overline{q}d$ and do not constrain these scenarios further.

\subsection{Collider searches}

Additional constraints on these scenarios come from collider searches of the {\emph{colored}} messengers $\Psi$. Here we focus on the minimal 4D models of Appendix~\ref{app:NB} (with $\xi\sim{\bf 3}$ of either $SU(3)_u$ or $SU(3)_d$). The warped 5D models introduced in the previous section have a more model-dependent collider phenomenology, and may or may not have light $\Psi$s, as argued in section~\ref{sec:stab}.

In the models of Appendix~\ref{app:NB} the messengers $\Psi$ have mass squared $m_*^2={m_\Psi^2+|\lambda\Sigma|^2}$, up to corrections of order $m_{t,b}^2/m_*^2\ll1$. They correspond to one or more families of $SU(2)_w$-singlet vector-like top or bottom quarks. At the LHC they are pair-produced by QCD interactions or singly-produced via the Yukawas, and are expected to decay into third generation SM quarks with ${\rm BR}(\Psi\to ht)\approx{\rm BR}(\Psi\to Z^0t)\approx\frac{1}{2}{\rm BR}(\Psi\to W^+b)\approx25\%$ for $\xi\sim{\bf 3}\in SU(3)_u$, whereas ${\rm BR}(\Psi\to hb)\approx{\rm BR}(\Psi\to Z^0b)\approx\frac{1}{2}{\rm BR}(\Psi\to W^-t)\approx25\%$ for $\xi\sim{\bf 3}\in SU(3)_d$. Current searches for heavy partners of the top and bottom quarks can be used to set an approximate bound $m_*\gtrsim700$ GeV~\cite{atlas}\cite{Aad:2014efa}\cite{Chatrchyan:2013uxa} on one-generation models.

\section{Discussion}
\label{sec:conclusions}

What makes the strong CP problem stand out from the list of puzzles in particle physics is the absence of an obvious ``environmental justification". While a small cosmological constant~\cite{Weinberg:1987dv}, the Planck-TeV hierarchy~\cite{Agrawal:1997gf}, the existence of dark matter and the rarity of anti-matter~\cite{Tegmark:2005dy} may be argued to be \emph{essential} to the existence of our universe, the smallness of the neutron EDM cries out for a dynamical explanation.

In this paper we discussed solutions to the strong CP problem based on spontaneous CP violation. The objective of these scenarios is to generate a CKM phase of order unity while simultaneously forbid large corrections to $\bar\theta$ after symmetry breaking. This poses severe constraints on the way in which the CP-odd scalars interact with quarks and, in sharp contrast to the QCD axion solution, also on the amount of quark flavor violation within the effective theory. 

In a Supersymmetric world, finding theories of this type would be much easier, because the SUSY non-renormalization theorems forbid perturbative contributions to $\bar\theta$~\cite{Ellis:1982tk}\cite{Hiller:2001qg} (up to  threshold effects). We can understand this result, and extend it to take into account non-perturbative contributions, using the spurion formalism of section~\ref{sec:spurion} and holomorphicity of the Wilsonian action~\cite{Seiberg:1993vc}: in a SUSY framework $\bar\theta$ is renormalized only if there exists a {\emph{holomorphic}}, CP-odd, flavor-singlet combination of the spurions.~\footnote{Thus, a flavor-singlet spurion, a sextet, a bi-fundamental, etc. would renormalize $\bar\theta$, whereas the choices $\xi\sim{\bf 3,8,}$ etc do not.} Unfortunately, SUSY is broken, and one should find a way to protect $\bar\theta$ without it.

From a general non-Supersymmetric perspective, the most favorable scenarios are those (see figure~\ref{fig}) in which the dominant flavor-violating parameters in the visible sector are (in a suitable field basis) the CP-even proto-Yukawa matrices $y_{u,d}$, accounting for the quark mass hierarchy, and the CP-odd spurion(s) $\xi$. The quark Yukawas and the topological angle are functions of these parameters, with $\xi\sim1$ to ensure a large CKM phase. We have seen that departures from the setup of fig.~\ref{fig} are very strongly constrained.

Our main result is a robust, model-independent criteria to solve the strong CP problem, see Section~\ref{sec:general}. Simply stated, one just needs to build a scenario of the type represented in fig.~\ref{fig} and make sure that there exists either a unique $\xi$ charged under the fundamental or the adjoint of $SU(3)_{u,d}$, or two CP-odd spurions, one in $SU(3)_u$ and one in $SU(3)_d$. If so, the accidental symmetries of the low energy theory guarantee that $\bar\theta$ first arises at sixth order in the SM Yukawas, comfortably below the bound (\ref{robust}), see table~\ref{table}. Solutions that do not meet these criteria usually require fine-tuning or non-trivial anti-correlations between the (a priori independent) flavor-violating parameters.

It is interesting to note that our conditions are neither stronger nor weaker than Barr's conditions~\cite{Barr:1984qx} for a vanishing {\emph{tree-level}} $\bar\theta$; they are just different. There exist models in Barr's class that do not meet our criteria (for instance the one originally proposed by Nelson~\cite{Nelson:1983zb}, that has two CP-odd spurions, one of which resides in $SU(3)_q$), and there are scenarios that satisfy our requirements but do not belong to Barr's class (such as the model of section~\ref{sec:5D} and some of those reviewed in Appendix~\ref{sec:app}).

Yet, the simplest 4D scenarios that satisfy our criteria are of the type presented in~\cite{Bento:1991ez}, which also fall in Barr's class (see Appendix~\ref{app:NB}). Current bounds on these scenarios are dominated by flavor and electro-weak data as well as LHC searches for top and bottom partners. These models provide an excellent example of well motivated new physics scenarios with potentially spectacular signatures at hadron colliders, but no obvious connection to the hierarchy problem.

In addition, a variety of novel solutions to the strong CP problem can be constructed in an extra dimensional setup. These 5D models of ``quark compositeness" find an obvious implementation within composite Higgs models, that are independently motivated by the hierarchy problem. We discussed in detail a realistic scenario with ``composite right-handed quarks", where CP violation is shined through the bulk by the scalar responsible for stabilizing the extra dimension. The model addresses both Higgs naturalness and the strong CP problem, and is compatible with a new physics scale above a few TeV. The strongest bounds here come from departures from the MFV paradigm (see fig.~\ref{figB}), as well as precision electro-weak constraints (which are basically insensitive to the new ingredients introduced to solve the strong CP problem). Scenarios with ``composite doublets" can be realized along similar lines.

We showed in sec.~\ref{sec:estimate} that the 4D and 5D models of App.~\ref{app:NB} and Sec.~\ref{sec:5D} belong to the same, larger class of CFTs reproducing the picture of fig.~\ref{fig}, but we have not been able to prove that the opposite is also true. It would be interesting to find models consistent with fig.~\ref{fig} which are not of the type described by eq.~(\ref{CFT}).

In the models depicted in figure~\ref{fig}, flavor-violation beyond the SM is controlled by $\xi$, and this points to interesting model-dependent correlations between the neutron EDM and flavor-violating observables, that can potentially be tested experimentally. For example, 5D models with $\xi\in SU(3)_d$ and $m_*$ in the multi-TeV range induce large flavor-violating effects in the down sector and a potentially testable neutron EDM --- from dimension-6 operators. On the other hand, if $\xi$ is charged under $SU(3)_u$ and $m_*$ sufficiently large, a $\bar\theta$ within the reach of future experiments may well be the only signature.

Finally, CP violation may leak in the lepton sector via a spurion carrying lepton flavor indices in a way analogous to the one described for quarks. Our results show that, as long as lepto-quark interactions are suppressed, CPV in the lepton sector is effectively unconstrained by the neutron EDM. Leptogenesis thus appears as a plausible candidate for the generation of a matter anti-matter asymmetry in these models.

%How plausible are these models compared to the QCD axion?

\acknowledgments

We would like to thank R. Sundrum and R. Mohapatra for useful discussions. This work was supported in part by NSF Grant No. PHY-1315155 and by the Maryland Center for Fundamental Physics.

\appendix

\section{4D models}
\label{app:NB}

The simplest models realizing the setup of fig.~\ref{fig} can be written in the notation of eq.(\ref{CFT}) as
\ba\label{minimalNB}
O_u=\overline{q}\widetilde{H}u,~~~~~~~O_d=\overline{q}Hd,~~~~~~~O=\overline{Q}\Psi~~~(Q=u~{\rm and/or}~d),
\ea
where $\Psi$ is a vector-like fermion with mass $m_\Psi$ and appropriate charges under the SM, and the CP-odd spurion is automatically in the fundamental of $SU(3)_{u,d}$. The case $Q=d, d_\Sigma=1$ (i.e. with $\Sigma$ a fundamental scalar) has been first presented in~\cite{Bento:1991ez}, see~\cite{Hiller:2001qg} for a realistic SUSY version. The models (\ref{minimalNB}) satisfy Barr's criteria~\cite{Barr:1984qx}, which appear here as a consequence of our requirements.

The scenarios (\ref{minimalNB}), including SUSY extensions, are appealing because very minimal. We emphasize however that
\ba\label{tune}
|Y_Q\lambda\langle\Sigma\rangle|\lesssim m_\Psi\lesssim|\lambda\langle\Sigma\rangle|
\ea
must be arranged to guarantee perturbativity. To see this note that a large mixing $\Psi-Q$ is needed to get $\theta_{\rm CKM}\sim1$, and this requires $m_\Psi\lesssim\lambda\langle\Sigma\rangle$. On the other hand, taking $m_\Psi/\lambda\langle\Sigma\rangle\ll1$ effectively enhances the coupling between the {\emph{physical}} heavy modes of mass $m_*^2=m_\Psi^2+|\lambda\Sigma|^2$ (a combination of $Q,\Psi$) and the SM weak doublet $q$. The latter coupling is $Y_QU{\lambda\langle\Sigma\rangle}/{m_\Psi}$, with $U$ a unitary matrix. Barring accidental cancellations among $y_{u,d}, \lambda\langle\Sigma\rangle$, requiring perturbativity below the cutoff thus sets an upper bound on $Y_Q{\lambda\langle\Sigma\rangle}/{m_\Psi}$ of order unity, see (\ref{tune}).

In models with $m_\Psi$ a bare mass defined at the UV cutoff $\Lambda$, the condition (\ref{tune}) is the result of a very peculiar choice of UV parameters. In fact, because the couplings $\lambda$ and $m_\Psi$ run very differently from the cutoff down to $\langle\Sigma\rangle\ll\Lambda$, from an effective field theory perspective (\ref{tune}) appears as a remarkable coincidence. This is even more evident if $\Sigma$ is a fermion bilinear. On the other hand, if for example 
$$
m_\Psi=\lambda'\langle S\rangle,
$$ 
for some scalar $S$ getting a vev {\emph{comparable}} to that of $\Sigma$, then (\ref{tune}) becomes $\lambda(\Lambda)\sim\lambda'(\Lambda)$, which is a perfectly legitimate and radiatively stable assumption on the short distance dynamics. Unfortunately, in these {\emph{natural}} models the minimality of (\ref{minimalNB}) is partially lost.

{In our warped model (\ref{tune}) is replaced by (\ref{scale}), which we argued is generic.}

\section{On the previous literature}
\label{sec:app}

%Besides the Nelson-Barr framework, the models share some analogy with the scenarios presented in~\cite{Hiller:2001qg} and~\cite{Harnik:2004su}.

\paragraph{HS}
\label{app:HS}
In SUSY models, the dominant contributions to the neutron EDM come below the SUSY-breaking scale, and the authors of ref.~\cite{Hiller:2001qg} showed that these can be made small provided (1) CP violation occurs at a scale $m_{\rm CP}$ much larger than SUSY breaking, (2) SUSY-breaking is communicated via some flavor-invariant mediator (e.g. gauge mediation), and (3) the SUSY-breaking sector interacts weakly with the CP-violating dynamics. Under these reasonable assumptions, the IR physics respects MFV, and contributions to $\bar\theta$ are utterly small~\cite{Ellis:1978hq}\cite{Dugan:1984qf}\cite{Khriplovich:1993pf}\cite{Hiller:2001qg}. This regime coincides with the limit of exact MFV, and trivially meets our criteria.

Ref.~\cite{Hiller:2001qg} generates the CKM phase either with the Nelson-Barr mechanism, which we discussed in Appendix~\ref{app:NB}, or by invoking a non-perturbative interaction between the SM quarks and messenger fields that directly couple to the CP-odd scalar. For the latter mechanism to work in the specific model discussed in~\cite{Hiller:2001qg}, the quark-messenger coupling must become strong exactly at the messenger mass scale, which is an independent relevant parameter. Besides bringing a calculability issue on the table, this poses a coincidence problem (mentioned in Appendix~\ref{app:NB}) that was not addressed in~\cite{Hiller:2001qg}. The present paper employs a similar mechanism, but solves these issues by constructing calculable (5D models) dual to scenarios with composite quarks.

\paragraph{HPSS} In the 5D model~\cite{Harnik:2004su}~\footnote{A realization on a warped background was presented in~\cite{Cheung:2007bu}.} NDA suggests that both scales $\lambda\Sigma, m_*$ are naturally set by the 5D cutoff, so the second condition in (\ref{robust}) is obtained naturally. This is analogous to what happens in the 5D example of sec.~\ref{sec:5D}.

Using a (dual) 4D language, we may view~\cite{Harnik:2004su}\cite{Cheung:2007bu} as a scenario in which CP is spontaneously broken at $O(1)$ by a strong dynamics (i.e. the bulk) that violates flavor. Our analysis suggests that in such a situation there is a priori no small parameter that suppresses $\bar\theta$. Indeed, in order to reduce unwanted UV contributions to $\bar\theta$, additional assumptions have been made in~\cite{Harnik:2004su}. The authors discuss 3 possibilities. First, the vev of the bulk CP-odd scalar may be much smaller than the cutoff --- this requires a fine-tuning. Second, the bulk CP-odd scalar may be assumed to have no overlap with the branes (by the bulk P invariance $\bar\theta$ can only appear on the boundaries), though no simple way to realize this possibility exists. Third, the CP-odd scalar may be promoted to a ${\bf 8}\in SU(3)_Q$, which is one of the possibilities discussed in table~\ref{table}. In this case $\bar\theta$ can be naturally below the current bounds. However, this would require a modification of the flavor structure of~\cite{Harnik:2004su}\cite{Cheung:2007bu}.

\section{A confining CP-violating sector}
\label{sec:conf}

Here we present a minimal CPV sector for fig.~\ref{fig}. The following discussion immediately applies to the class of models captured by (\ref{CFT}).

An elegant way to achieve $m_{\rm CP}\ll\Lambda$ is to introduce chiral fermions $Q$ charged under a confining strong dynamics and a weakly gauged $U(1)_\xi$. A minimal field content shown in the table \ref{minimal}. This is basically a copy of 2-flavor QCD, with $U(1)_\xi$ generated by $\sigma^3$ of $SU(2)_L$, and
\ba
\Sigma_{ij}=\overline{Q_{jR}}Q_{iL}.
\ea
The gauged $SU(n)$ gets strong at a scale $m_{\rm CP}\ll\Lambda$ and produces 3 Goldstone modes, one combination of which is eaten by the $U(1)_\xi$ vector. We can perform gauge rotations such that the physical Nambu-Goldstone modes are the two angles $\pi_{1,2}$ defined by
\ba
\Sigma\propto\left( \begin{array}{cc}  
e^{i\pi_2}\cos\pi_1 & \sin\pi_1 \\
-\sin\pi_1 & e^{-i\pi_2}\cos\pi_1
\end{array}\right).
\ea
In this basis CP is broken if $\pi_2\neq0$.

%%%%%%%%%%%%%%%%%%%%%
\begin{table}[t]%[thh]
\begin{center}
\begin{tabular}{c|cc} 
\rule{0pt}{1.2em}%
$$ &  $SU(n)$ & $U(1)_\xi$ \\
\hline
%\hline\hline
%\hline
$Q_{iL}$ & ${\bf n}$ & ${\bf \pm\frac{1}{2}}$\\
$Q_{jR}$ & ${\bf n}$ & ${\bf 0}$\\
\end{tabular}
\end{center}
\caption{\small Minimal CP-violating sector, with $i,j=1,2$.
\label{minimal}}
\end{table}
%%%%%%%%%%%%%%%%%%%%%%%

The global $SU(2)_L\times SU(2)_R$ is explicitly broken by the gauged $U(1)_\xi$, the irrelevant coupling $\lambda\Sigma O$, and cutoff-suppressed $QQQQ$ operators. All of these effects contribute to a potential for $\Sigma$. The gauge coupling gives a positive mass squared to $\pi_{1,2}$ and tends to align the vacuum along the direction $\Sigma\propto1$. For ${g_\xi^2}/{16\pi^2}<{m_{\rm CP}^2}/{\Lambda^2}$, which is a reasonable condition in a realistic model (see below), the potential is dominated by the 4-fermions operators. For generic (real) coefficients of $QQQQ$ one finds that CP is spontaneously broken and $\pi_{1,2}\sim1$.~\footnote{Analogously, QCD spontaneously breaks CP~\cite{Dashen:1970et} for generic quark masses.} This generates a complex tadpole $\lambda\langle\Sigma\rangle$, see~(\ref{phiaction}).

Let us now verify that all contributions to $\bar\theta$ from physics $\gtrsim m_{\rm CP}$ are small. In fact, these can be naturally suppressed if CP violation is soft, i.e. $\langle\Sigma\rangle\sim m_{\rm CP}^{d_\Sigma}$ with $m_{\rm CP}\ll\Lambda\sim10^{18}$ GeV. In particular, effects controlled by $\Sigma^\dagger\Sigma$ (such as a bare $\bar\theta$) are negligible as long as
\ba\label{stronger}
\left(\frac{m_{\rm CP}}{\Lambda}\right)^{d_\Sigma}<10^{-5},
\ea
with $d_\Sigma$ the scaling dimension of $\Sigma$. For this particular CPV sector $d_\Sigma\simeq3$, and it is sufficient to require $m_{\rm CP}\lesssim10^{-2}\Lambda$. The above discussion suggests that CP will generically be spontaneously broken as long as $g_\xi\lesssim0.1$.

Below the scale $m_{\rm CP}$ the CP-violating sector is described by the Goldstone modes. Loops of $\pi_{1,2}$ could in principle give a dangerous correction to $\bar\theta$, but these can be made parametrically small even if $\lambda\sim1$ by taking ${d_\Sigma+d_O}>4$, with $d_O$. Under these conditions, CP violation at scales $<m_{\rm CP}$ is controlled by the complex matrix $\xi\propto J$, realizing the framework of fig.~\ref{fig}.

A final comment is in order. The CP-violating dynamics has hadrons at the scale $m_{\rm CP}$, as well as Goldstone modes of mass $m_{\pi_{1,2}}\sim m_{\rm CP}^2/\Lambda\ll m_{\rm CP}$. These are all very weakly coupled to ordinary matter and could represent a cosmological hazard. We avoid this problem by demanding that $m_{\pi_{1,2}}\sim10^{14}\left(\frac{\Lambda}{10^{18}~{\rm GeV}}\right)> T_{\rm RH}$, where $T_{\rm RH}$ is the re-heating temperature. Under this hypothesis the population of the exotic states will be washed out by inflation. Importantly, this assumption also ensures that possible domain walls generated at spontaneous CP breaking at $T_{\rm CP}\sim m_{\pi_{1,2}}$ do not alter the standard big-bang cosmology. %Baryogenesis may be achieved via standard leptogenesis after re-heating. 

%%%%%%%%%%%%%%%%%%%%%%%%%%%%%%%%%%%%%

 \end{document}